\documentclass[prb,showpacs,byrevtex,twocolumn]{revtex4}
\usepackage{graphicx}
\usepackage{float}
\usepackage{bm}
\begin{document}

\bibliographystyle{apsrev}

\preprint{Draft version, not for distribution}

%
%

\title[coherence]{Predicting new superconductors and their critical temperatures using unsupervised machine learning}

%
%
%
\author{B. Roter}
\affiliation{Department of Physics, The University of Akron,
Akron, Ohio 44325, USA}%
\author{S.V. Dordevic}
\email{dsasa@uakron.edu}%
\affiliation{Department of Physics, The University of Akron,
Akron, Ohio 44325, USA}%

\date{\today}

%
%
\begin{abstract} We used the superconductors in the SuperCon database
to construct element vectors and then perform unsupervised learning
of their critical temperatures (T$_c$). Only the chemical composition of
superconductors was used in this procedure. No physical predictors
(neither experimental nor computational) of any kind were used. We
achieved the coefficient of determination R$^2$~$\simeq$~0.93,
which is comparable and in some cases higher
then similar estimates using other artificial intelligence
techniques. Based on this machine learning model, we predicted
several new superconductors with high critical temperatures.
We also discuss the factors that limit the learning process and
suggest possible ways to overcome them.
\end{abstract}

%
%
%


\maketitle

\section{Introduction}

Quantum supremacy was recently achieved by Google using a superconducting
microprocessor Sycamore \cite{arute19}. An avalanche of similar results is
now expected. The future of superconductors has never looked brighter.
However, if Sycamore and other superconducting microprocessors are to
find a wider circle of users, their operating temperature will have to
be increased significantly. Sycamore is made of aluminum (T$_c$ = 1.175 K)
and indium (T$_c$ = 3.41 K) and operates at temperatures below 20 mK
\cite{arute19}. Such low temperatures require a dilution refrigerator
with $^3$He, which is exceedingly
rare and expensive. This clearly illustrates the need for new
superconducting materials with higher critical temperatures.
However, finding new superconductors, especially with
high T$_c$, is a very difficult endeavor \cite{uchida-book}.

In recent years, there has been a surge of interest in using artificial
intelligence (AI), in particular machine learning (ML) and deep learning 
(DL), in materials physics \cite{isayev15,hill16,yuan19}.
The idea is that by using the existing information in materials' databases,
one can predict new materials with certain desired properties.
In particular, several
attempts have been made in predicting the critical temperatures
of superconductors, or more generally, predicting new materials with
potentially high T$_c$. Several prominent efforts have been by Stanev et.al
\cite{stanev18}, Hamidieh \cite{hamidieh18} and Zeng et al. \cite{zeng19}.
Different AI approaches were used in these papers: Stanev et al. used
both classification and regression models, Hamidieh used an XGBoosted statistical
model, and Zeng et al. used convolutional neural networks (CNN).

In a recent work by Zhou et.al \cite{zhou18}, the properties
of the atoms were learned from the chemical compositions of compounds
from a large database, without any additional information. Inspired by
this approach, we made a similar attempt in predicting new superconducting
materials and their critical temperatures. The {\it only} predictor
used is the chemical composition of compounds (both superconducting
and non-superconducting), which is readily available in the existing
databases and does not require any post-processing.
We employed what Zhou et.al \cite{zhou18} called
unsupervised machine learning and achieved statistical parameters
comparable, and in some instances exceeding previous attempts.
Below we describe in details the procedure used, and then the
results of our study. We also discuss the factors that limit the
learning process, most notably the wrong entries into the database.

\section{SuperCon database}

SuperCon is currently the biggest and most comprehensive database
of superconductors in the world \cite{supercon}.
It is free and open to the public, and it has been used
in almost all AI studies of superconductors \cite{stanev18,hamidieh18,zeng19}.
At the time when we downloaded it,
it contained almost 34,000 entries. Fewer than 100 of them had
errors in their chemical formulas and were removed.
About 7,000 entries did not have the values of 
T$_c$ reported and were also removed. The remaining 27,000 were
used for our ML calculations. However, fewer then 100 entries
had T$_c$~=~0. The importance of non-superconducting compounds
for AI calculations was noted previously \cite{zeng19}.
In order to improve the predictive power of ML models,
we supplemented SuperCon database with about 3,000 non-superconducting
compounds, mostly insulators, semiconductors and some non-superconducting
metals and alloys. In total, our database had about 30,000 entries.
There were several thousands of multiple entries which were all
kept in order to improve their statistical significance \cite{zhou18}.
The entire database was used as the training set for both classification
and regression models discussed in Sections V and VI below.

\section{Element-vectors}

Once the database was created, we parsed all the formulas and wrote
the chemical content for each superconductor into a matrix which we call
the chemical composition matrix. The corresponding values of T$_c$
were written in a separate vector. The matrix has about 30,000
rows and 96 columns. The number of columns is determined by
the elements present in the chemical formulas. The heavies
element that appears in any superconductor in the database is
Curium (Z~=~96) and consequently the chemical composition matrix
has 96 columns. We also note that the matrix is extremely sparse, 
as more than 96~$\%$ of its elements are zeros.

\begin{figure*}[tbp]
\vspace*{0.0cm}%
\centerline{\includegraphics[width=6in]{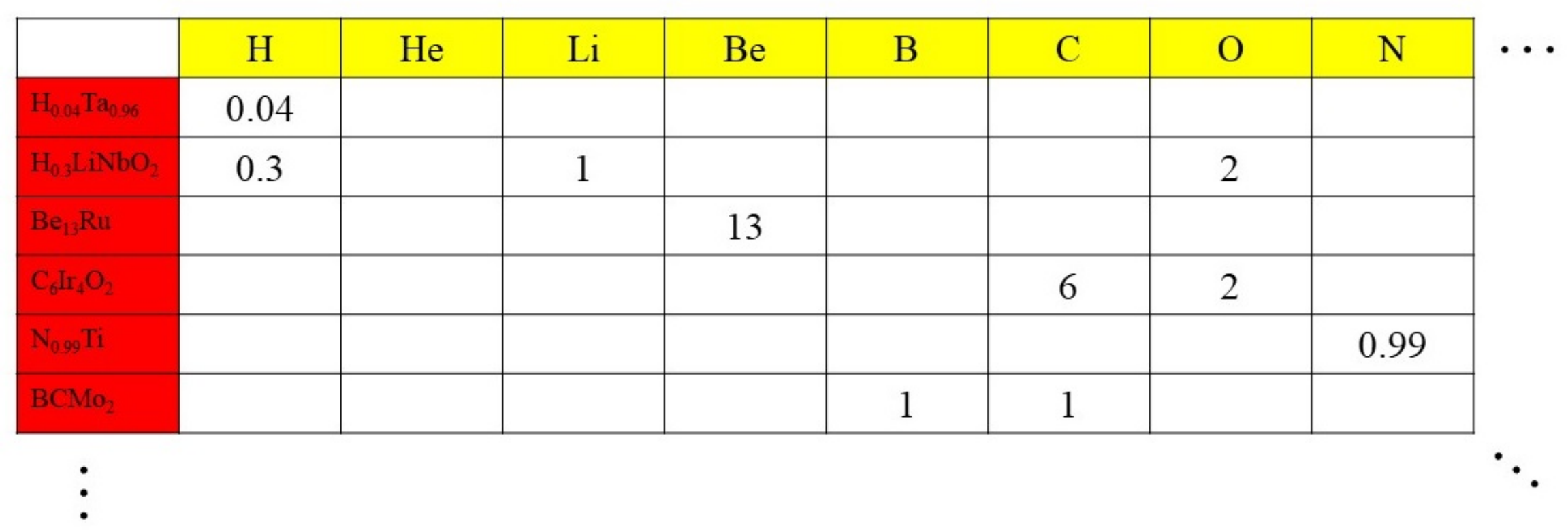}}%
\vspace*{-0.25cm}%
\caption{(Color online). Upper left corner of the chemical
composition matrix. The size of the whole matrix is (approximately)
30,000 $\times$ 96: (approximately) 30,000 entries in the database
and 96 elements. The columns of the matrix represent element-vectors
that we used as the {\it only} predictors in our calculations.}
\vspace*{0.0cm}%
\label{fig:chem-comp}
\end{figure*}

In Figure~\ref{fig:chem-comp} we show a very small portion
(upper left corner) of the chemical composition matrix. By
analogy with Ref.~\onlinecite{zhou18} we call the columns of
the matrix element-vectors (or atom-vectors). They contain
the information about the superconductors in the database
and are the {\it only} predictors used in training the models.


\section{Unsupervised machine learning}

In this section we describe what was previously referred to as
unsupervised machine learning \cite{zhou18}. At the heart of this
approach is the so-called Singular Value Decomposition (SVD), or
equivalently Principle Component Analysis (PCA) \cite{wall-book}. 
Once the chemical composition matrix is formed as described above 
(Fig.~\ref{fig:chem-comp}), it is decomposed according to \cite{wall-book}
\begin{equation}
  X =  U S V^T
  \label{eq:svd}%
\end{equation}
where U and V are unitary matrices, and S is a diagonal
matrix whose values are called singular values.
The columns of matrix U will be referred as element-eigenvectors
(or atom-eigenvectors) by analogy with Ref.~\onlinecite{alter00}.
These element-eigenvectors contain higher dimensional information
about the chemical composition of superconductors, and they were
used as the {\it only} predictors by ML models. The rank of the chemical composition
matrix described in the previous section (Fig.~\ref{fig:chem-comp})
is 83, which indicates that
one can use any number of element-eigenvectors up to 83. Our
calculations indicate that with as few as 10 element-eigenvectors one
can achieve significant improvements over the calculations using
raw element-vectors described in the previous section.

\section{Classification models}

To check the effectiveness of the procedure described above,
we first constructed classification models using only 
element-eigenvectors as predictors.  These classification 
models were designed for predicting whether a compound 
is a superconductor or not. A myriad of 
different training algorithms were tested, such as the Bagged 
Tree, Boosted Tree and Gaussian Support Vector Machines. After 
a number of tests, we concluded that the 
method called k-Nearest Neighbors (KNN) was the most 
consistently accurate for this particular task.

The KNN approach assumes that 
anything similar exist within close proximity of each other, and 
it depends on that assumption being true enough for the algorithm 
itself to be useful.  In other words, KNN is designed to exploit
the ideas of similarity and mathematical distances.  Given a set 
of points and a distance function, KNN allows one to find the k 
closest points in that set to either a point or collection of 
points of interest, irrespective of any labeling.  The latter 
group of points represent classes, and those classes, in our 
case, represent whether a chemical compound is non-superconducting 
or superconducting.

\begin{figure}[t]
\vspace*{0.0cm}%
\centerline{\includegraphics[width=3.5in]{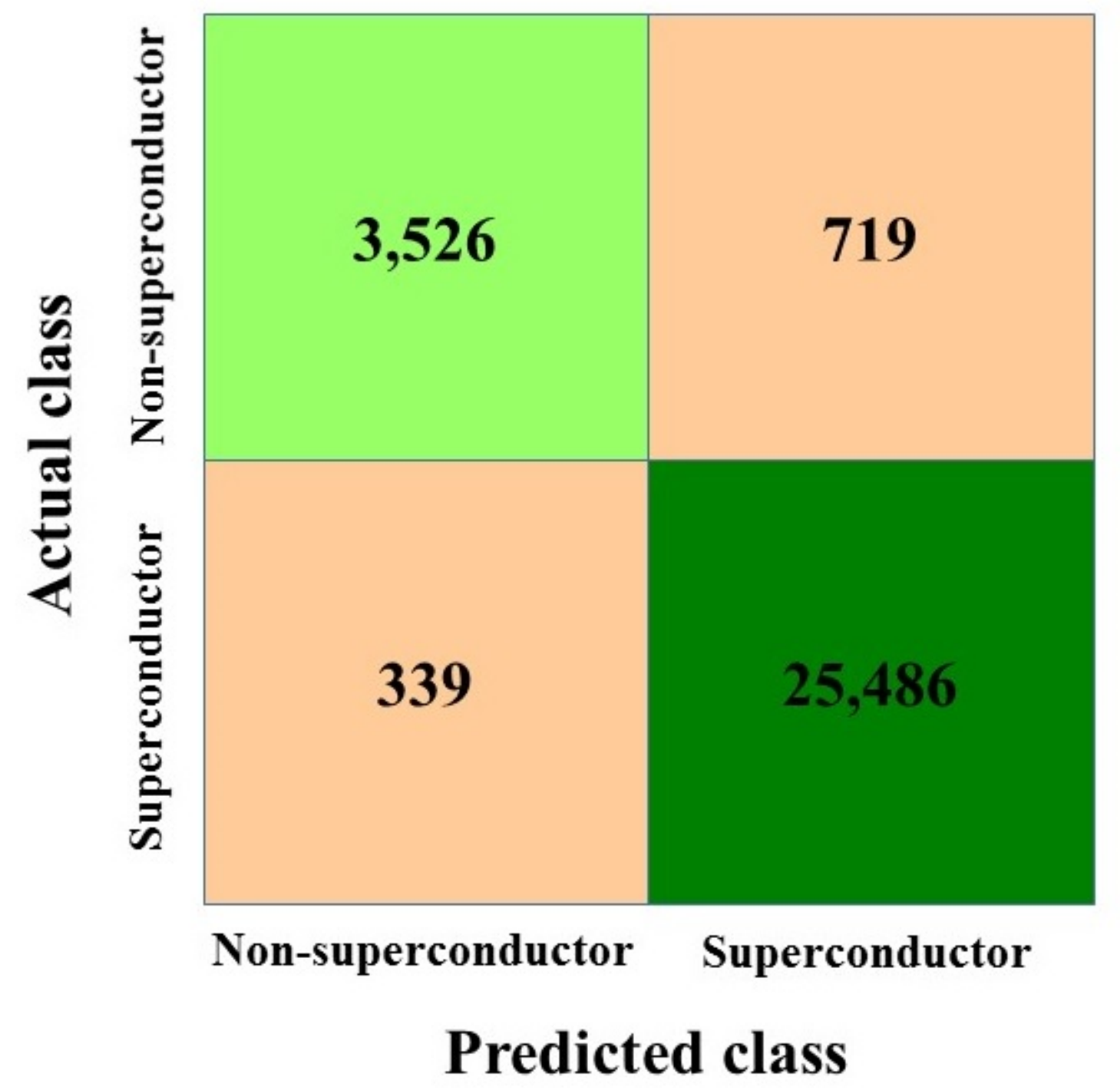}}%
\vspace*{0.0cm}%
\caption{(Color online). A confusion matrix for the 
classification model. The overall accuracy is 96.5~$\%$. }
\vspace*{0.0cm}%
\label{fig:class}
\end{figure}

Applying the KNN method to our dimension-reduced chemical composition 
matrix U (Eq.~\ref{eq:svd}), we were able to create a classification model that was 
96.5~$\%$ accurate. This value exceeds those obtained by other AI 
techniques \cite{stanev18,zeng19}. Figure~\ref{fig:class} shows 
the corresponding confusion matrix, which illustrates the need 
for including even more non-superconducting entries into the database.


\section{Regression Models}

One can construct the regression models using the raw 
element-vectors described above in Section III (Fig.~\ref{fig:chem-comp}). 
The values of statistical parameters \cite{stat-param}
achieved using these element-vectors are: the coefficient of
determination R$^2$~$\simeq$~0.90 and the root-mean-square
error RMSE~$\simeq$~9.67~K. As good as these numbers are, they can be 
further improved with the help of SVD (Eq.~\ref{eq:svd}).
Using element-eigenvectors as the only predictors, we also constructed 
our regression models. A number of different training methods were then 
performed on them, such as Exponential Gaussian Process Elimination, Fine Tree, 
Boosted Tree, as well as a Gaussian Support Vector Machine (SVM). 
Of these methods, however, 
an algorithm known as the Bagged Tree was the most accurate on a consistent 
basis in predicting the values of T$_c$ for our input compounds.

The Bagged Tree method is a variant of Random Forests algorithms \cite{breiman01}. 
It combines multiple decision trees in order to output 
better predictions - a stark contrast from just creating one decision tree. 
The underlying theory for this technique is that multiple weak learners 
should be able to combine into a much more robust form.  Typically, 
with ML algorithms involving decision trees, altering the training 
data in any way can yield completely different trees, thus yielding 
completely different predictions. The method of bagging is designed 
to help greatly mitigate the high-variance nature of these decision 
trees.  The only parameter associated with this technique is the 
size of the training data, corresponding to the number of trees to 
include in the bagging process.

\begin{figure}[t]
\vspace*{0.0cm}%
\centerline{\includegraphics[width=3.5in]{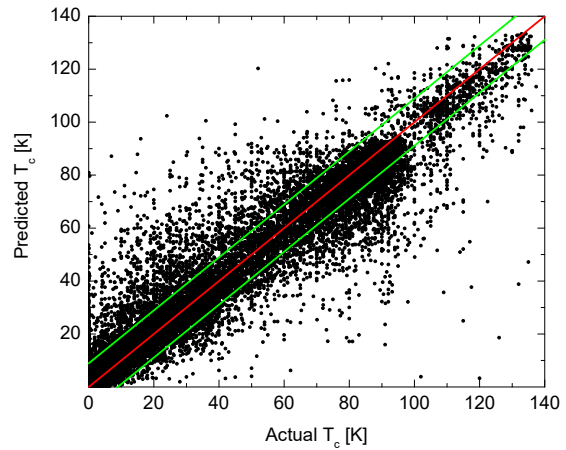}}%
\vspace*{0.0cm}%
\caption{(Color online). A plot of predicted T$_c$ versus
actual T$_c$. The achieved statistical parameters are
R$^2$~$\simeq$~0.93 and RMSE~$\simeq$~8.91~K.}
\vspace*{0.0cm}%
\label{fig:predicted}
\end{figure}

Applying the Bagged Tree algorithm to our dimension-reduced chemical 
composition matrix, we were able to create a regression model that 
achieved R$^2$~$\simeq$~0.93 and RMSE~$\simeq$~8.91~K. These values 
of R$^2$ and RMSE are comparable and in some cases higher than the values Stanev et al., 
Hamidieh, and Zeng et al. achieved using their respective training 
algorithms. In Figure~\ref{fig:predicted} below we display the results of our calculations. 
The values of predicted T$_c$ are plotted as a function of actual 
T$_c$ for all 30,000 compounds in the SuperCon 
database. We notice that plot does not reveal any systematic 
offsets, and vast majority of the data points are clustered 
around a line with unit slope (red line) in the $\pm$8.91~K range (green lines). 
The outliers that appear in Figure~\ref{fig:predicted} 
could be due to incorrect initial values of T$_c$, which are 
discussed in more detail in Section VIII below.


\section{Predictions}

Using our best model (Fig.~\ref{fig:predicted}) we also
make predictions of new superconducting materials. For that purpose we
downloaded the entire Crystallography Open Database (COD) \cite{cod}, which
contains about 37,000 inorganic compounds and alloys from which we
make predictions. Not surprisingly, most compounds predicted to have 
a relatively high T$_c$ are oxides \cite{stanev18}. Some of the them 
are listed in Table~\ref{pred}. Interestingly, our calculations
are predict a number of non-oxide and non-iron based materials
with T$_c$ in the range 40--60~K. Some of the most promising 
examples are also listed in Table~\ref{pred} \cite{full-list}.

\begin{table}[t]
\centering
\begin{tabular}{|l|c|c|}
\hline

Compound & Predicted T$_c$ (K) & Comments  \\ \hline \hline
AlBaCaF$_7$                &   46  &  \\ 
As$_4$BaCu$_8$             &   50  &  \\ 
BaCu$_4$S$_3$              &   31  &  \\ 
CrCuSe$_2$                 &   26  &  \\ 
LiRbS                      &   21  &  \\ 
AlB$_4$Cr$_3$              &   50  &  \\ 
AlBa$_3$P$_3$              &   38  &  \\ 
BaCuTe$_2$O$_7$            &   54  &   see Ref.~\onlinecite{zeng19}  \\  
BaCu$_3$Br$_2$O$_4$        &   60  &   see Ref.~\onlinecite{zeng19}  \\  
Ba$_5$Br$_2$Ru$_2$O$_9$    &   57  &   see Ref.~\onlinecite{stanev18} \\  
CaCu$_2$Eu$_2$O$_6$        &   65  &   \\
Cr$_2$CuO$_4$              &   50  &    \\
Cu$_3$Na$_7$O$_8$          &   67  &    \\
Cl$_2$Sr$_2$CuO$_2$        &   27  &    \\     \hline             
\end{tabular}
\caption{Compounds from COD database \cite{cod} predicted to be 
superconducting \cite{full-list}. In addition to a number of oxide materials, our 
models also predicted several non-oxide materials with relatively high T$_c$.}
\label{pred}
\end{table}

We also included in this analysis the materials previously predicted to be
superconducting by Stanev et.al. \cite{stanev18} (without T$_c$) and Zeng
et.al. \cite{zeng19}. We find some of them to be superconducting with
almost the same T$_c$. For example, Zeng
et.al. \cite{zeng19} predicted BaCu$_3$Br$_2$O$_4$ to be superconducting
with T$_c$ of 60 K, which is the same T$_c$ that we predicted. 
On the other hand, Zeng et.al. found Na$_3$(TiS$_2$)$_{10}$ to be superconducting 
with T$_c$= 40.71~K, which we found to be non-superconducting. 
This illustrates intrinsic fragileness of AI methods.

\section{Limiting factors}

\begin{figure}[t]
\vspace*{-1.0cm}%
\centerline{\includegraphics[width=3.5in]{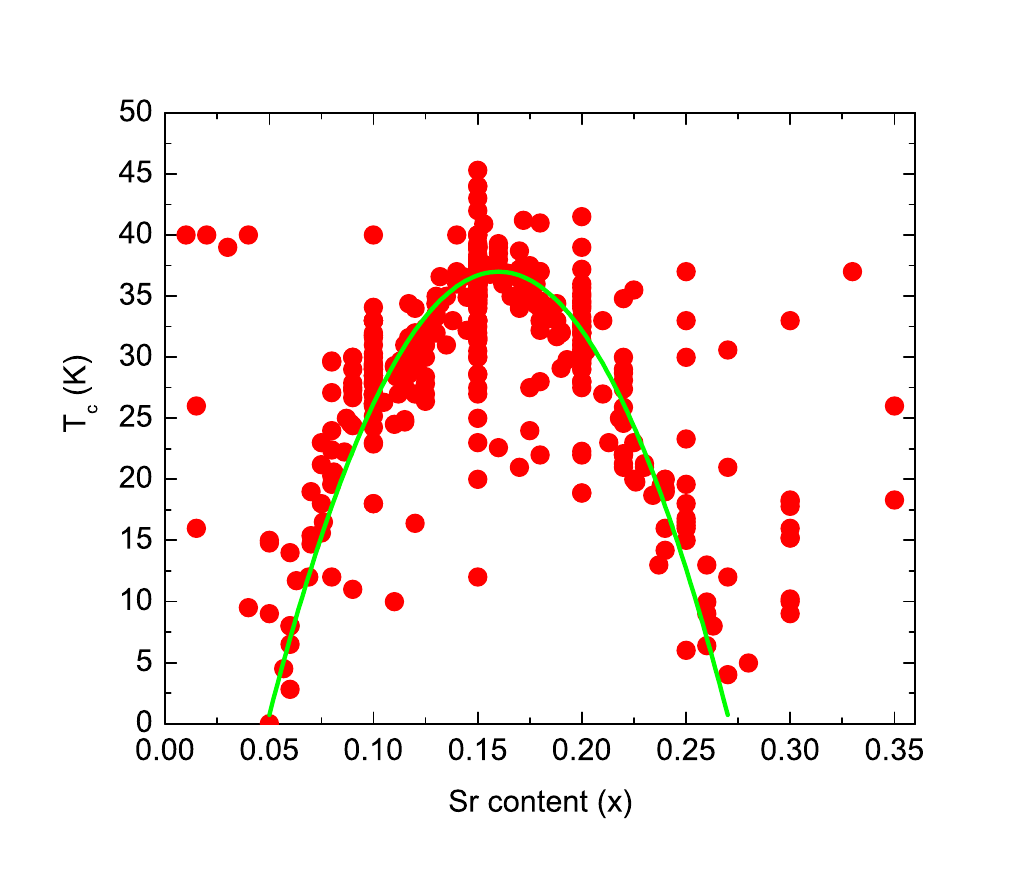}}%
\vspace*{-0.5cm}%
\caption{(Color online). A plot of T$_c$ versus strontium doping (x) 
for all 550 La$_{2-x}$Sr$_x$CuO$_4$ superconductors from the SuperCon database. Green line
is the expected T$_c$(x) behavior for this family of cuprates \cite{tallon01}.}
\vspace*{0.0cm}%
\label{fig:lsco}
\end{figure}

As the most important limiting factor of ML we identified the wrong entries into
the database. To illustrate that point in Fig.~\ref{fig:lsco}
we display the doping (x) dependence of T$_c$ for all La$_{2-x}$Sr$_x$CuO$_4$
(LSCO) superconductors in the SuperCon database. There are a total of 550 entries,
which are represented by red points. The green curve represents the expected doping
dependence of T$_c$ for this family of cuprate superconductors \cite{tallon01}. 
As can be seen from the plot, there are
a number of outliers. In addition, for many doping levels there are a
large number of points located either below or above the expected value.
Similar wrong entries are found in other cuprate families, 
as well as in iron-based superconductors. We estimate that there could be 
as many as 20~$\%$ of wrong entries in the entire database.
Our calculations reveal that when they are removed from the database,
the values of the statistical parameters can be further improved.



\section{Summary}

In summary, we used unsupervised machine learning to make predictions of
new compounds that are possibly superconducting and their
critical temperatures T$_c$. Using {\it only} the chemical composition of
superconductors from the SuperCon database we created a number
of models and achieved R$^2$~$\simeq$~0.93 and RMSE~$\simeq$~8.91~K.
These statistical parameters are comparable and in some cases exceed those
obtained with other AI studies that used a number of physical predictors
(both experimental and computational).

Our results indicate that one does not need predictors such as
the number of valence electrons, electronegativity, covalent radius, electron 
affinity or the number of unfilled orbits to achieve significant predictive power.
We argue that those predictors are not directly relevant for superconductivity
and that is the reason they did not lead to any significant improvements
of statistical parameters (R$^2$ and RMSE). We suggest
that physical predictors more closely related to superconductivity
should be used. Those include, for example, normal state resistivity
(or conductivity), superfluid density (or penetration depth),
band structure features, Fermi energy, etc. They have been shown to be closely 
related to T$_c$ \cite{homes04,dordevic13,liu18,uemura19} and,
in our opinion, would lead to improved models. Unfortunately there is
currently no comprehensive database that contains the values of these
parameters for a large number of superconductors. SuperCon does have some
of them, such as the penetration depth, critical fields, energy gap, etc.
However, the number of entries with these values reported is too small
for any meaningful AI calculations. For example, the values of the penetration
depth or the energy gap are reported for fewer than 1,200  superconductors
in the SuperCon.

We also showed that the limiting factor for achieving higher predicting
power is the quality of entries in the SuperCon database. If the number
of wrong entries can be reduced in the future, for example by human curation,
then the predictive power of ML models will inevitably be improved. 

Using our best models,
we also made predictions of new superconductors and their T$_c$'s. Running
the models on inorganic compounds from the COD database we have identified
a number of materials that are potentially superconducting, with relatively
high T$_c$. Future transport and/or thermodynamic measurements will determine 
how accurate any of these ML predictions \cite{stanev18,hamidieh18,zeng19} are.



%
%

\begin{references}


\bibitem{arute19} F. Arute, et al., Nature {\bf 574}, 505 (2019).

\bibitem{uchida-book} S. Uchida, {\it High Temperature Superconductivity: 
The Road to Higher Critical Temperature}, Springer Japan (2015). 

\bibitem{isayev15} O. Isayev, D. Fourches, E.N. Muratov, C. Oses, K. Rasch,
A. Tropsha and S. Curtarolo, Chemistry of Materials {\bf 27}, 735 (2015). 

\bibitem{hill16} J. Hill, G. Mulholland, K. Persson, R. Seshadri, C. Wolverton
and B. Meredig, MRS Bulletin, {\bf 41}, 399 (2016). 

\bibitem{yuan19} J. Yuan, V. Stanev, C. Gao, I. Takeuchi and K. Jin,
Superconductor Science and Technology, {\bf 32}, 123001 (2019). 

\bibitem{stanev18} V. Stanev, C. Oses, A.G. Kusne, E. Rodriguez, J. Paglione,
S. Curtarolo and I. Takeuchi, NPJ Computational Materials, {\bf 4}, 29 (2018).

\bibitem{hamidieh18} K. Hamidieh, Computational Materials Science, {\bf 154},
346 (2018).

\bibitem{zeng19}  S. Zeng, Y. Zhao, G Li, R. Wang, X. Wang and J. Ni,
NPJ Computational Materials, {\bf 5}, 84 (2019).   

\bibitem{zhou18} Q. Zhou, P. Tang, S. Liu, J. Pan, Q. Yan and S.-C. Zhang,
Proceedings of the National Academy of Sciences,  {\bf 115}, E6411 (2018).  

\bibitem{supercon} SuperCon database: https://supercon.nims.go.jp

\bibitem{wall-book} M.E. Wall, A. Rechtsteiner and L.M. Rocha, in: {\it A Practical
Approach to Microarray Data Analysis}, D.P. Berrar, W. Dubitzky and M. Granzow eds.,
pp. 91-109, Kluwer: Norwell, MA (2003) .   

\bibitem{alter00} O. Alter, P.O. Brown and D. Botstein,
Proceedings of the National Academy of Sciences, {\bf 97}, 10101 (2000). 

\bibitem{stat-param} R$^2$ and RMSE are conventionally defined 
\cite{bishop-book,friedman-book} as R$^2$ = 1 - SSE/SST
(SSE is the sum of squared error and SST is the sum of squared total)
and RMSE = $\sqrt{\sum_{i=1}^N (T_{c,predicted} - T_{c,actual})^2/N}$.

\bibitem{bishop-book} C.M. Bishop, {\it Pattern Recognition and Machine Learning},
Springer-Verlag New York (2006).

\bibitem{friedman-book} T. Hastie, R. Tibshirani and J. Friedman,
{\it The Elements of Statistical Learning}, Springer New York (2001).

\bibitem{breiman01} L. Breiman, Machine Learning {\bf 45}, 5 (2001). 

\bibitem{cod} Crystallography Open Database: https://www.crystallography.net

\bibitem{full-list} Full list of predictions is available upon request. 

\bibitem{tallon01} J.L. Tallon and J.W. Loram, Physica C {\bf 349},
53 (2001).  

\bibitem{homes04} C.C. Homes, S.V. Dordevic, M. Strongin, D.A. Bonn,
R. Liang, W.N. Hardy, S. Komiya, Y. Ando, G. Yu, N. Kaneko,
X. Zhao, M. Greven, D.N. Basov and T. Timusk, Nature {\bf 430}, 539 (2004).

\bibitem{dordevic13} S.V. Dordevic, D.N. Basov and C.C. Homes, Scientific 
Reports {\bf 3}, 1713 (2013). 

\bibitem{liu18} Y. Liu, N. Chen and Y. Li, arXiv:1811.12171.   

\bibitem{uemura19} Y.J. Uemura, Physical Review Materials {\bf 3}, 104801 (2019). 


\end{references}

\end{document}